\documentclass[fleqn,12pt,twoside]{article}
\usepackage{espcrc1}

\newcommand{\AmS}{{\protect\the\textfont2
  A\kern-.1667em\lower.5ex\hbox{M}\kern-.125emS}}

\title{Reductions of the Volterra lattice}

\author{A.K. Svinin\address[ISDCT]{Institute of System Dynamics and Control
        Theory, Siberian Branch of Russian Academy of Sciences, 
        P.O. Box 1233, 664033, Irkutsk, Russia}%
        \thanks{This research was supported by INTAS grant 2000-15 and
        RFBR grant 03-01-00102.}}

\begin{document}

% typeset front matter
\maketitle

\begin{abstract}
We exhibit three classes of algebraic constraints which are shown
compatible with Volterra lattice.
\end{abstract}

\section{Introduction}

In this Letter we discuss classical discrete system --- Volterra lattice \cite{volterra}
\begin{equation}
\frac{\partial r(i)}{\partial t} = r(i)\left(r(i+1)-r(i-1)\right).
\label{VL}
\end{equation}
Physical applications of this differential-difference system are well-known
(see, for example, Refs. \cite{zakharov1}, \cite{zakharov2}).
In particular, the system (\ref{VL}) can be interpreted as kinetic equation
describing stimulated scattering of plasma oscillations by ions.
This system
has been thoroughly studied for a number of initial conditions
\cite{manakov}, \cite{veselov}, \cite{ver} by inverse scattering transform
method. A number of works is concerned with the question: what a boundary 
conditions are consistent with higher flows in Volterra and Toda 
lattice hierarchy (see, for example \cite{bog}, \cite{gurel}, \cite{h1}, 
\cite{h2}, \cite{dam1}, \cite{dam2}). In particular some of their results show
that imposing some special boundary conditions for the lattices yields 
finite-dimensional systems corresponding to finite growth Lie algebras.  

Our principal goal in the Letter is to show three denumerable
classes of invariant submanifolds of the Volterra lattice which are defined
by some algebraic constraints. We show that each of these constraints are 
compatible with the Volterra equation itself and do not analyse their 
compatibility with the higher flows. As a result we are forced to consider 
finite-dimensional systems of ordinary differentional equations with rational
dependence on unknown functions suplemented by discrete symmetry 
transformation.

To make the matter more clear,
let us remind firstly the notion of differential constraints compatible with
a given system of differential equations \cite{sidorov}.  For our aims, we may
restrict ourselves by consideration of scalar evolutionary equation $E$ in
the following form:
\begin{equation}
\frac{\partial r}{\partial t} = F(r, r^{\prime},..., r^{(m)}),\;\;
\phantom{q}^{\prime}\equiv\frac{\partial}{\partial x}.
\label{E}
\end{equation}
Let us denote by $[E]$ the union this equation and its differential
consequences with respect to $x\in{\bf R}^1$. Let the equation (\ref{E}) be
supplemented by differential constraint $H$
\begin{equation}
h(r, r^{\prime},..., r^{(n)})=0,
\label{H}
\end{equation}
where $h$, as well as $F$, is supposed to be some locally analytic function of its arguments.
One says that differential constraint (\ref{H}) is compatible with (\ref{E})
or, in other words, define invariant submanifold for (\ref{E}), if
\begin{equation}
D_t(h)|_{[E]\cap[H]}=0,
\label{det}
\end{equation}
where $D_t$ stands for total derivative with respect to $t\in{\bf R}^1$.
The equation (\ref{det}), whose solutions are some differential functions, is
reffered to as determining one \cite{sidorov}.

The situation is considerably simplified when one can resolve (\ref{H}) with
respect to higher-order derivative as
\[
r^{(n)}=S(r,r^{\prime},..., r^{(n-1)}).
\]
Then practical recipe to solve determining equation (\ref{det}) consists of
successively replacing $r^{(n)}\rightarrow S$.

The method of differential constraints allows us to select some classes of
partial solutions of given equation (system of equations) by solving (\ref{H})
and further analysis. Observe that this method can be applied to both
integrable and nonintegrable equations. For integrable equations one can
expect that differential constraint being considered as ordinary differential
equation turns out to be integrable in some sense.

Let us return now to the Volterra lattice which is evolutionary equation of the
form
\[
\frac{\partial r(i)}{\partial t} = F(r(i+1), r(i), r(i-1)).
\]
Here the discrete variable $i\in{\bf Z}$ plays the role of ``space" variable
$x$ in (\ref{E}). Additional constraints in this case are not differential
but algebraic and solutions of determining equation are some locally analytic
functions
\[
h=h(r(i),..., r(i+n)).
\]
We believe that determining equation (\ref{det}) can be successfully applied
in discrete case and show this on example of Volterra and Toda lattice.

In Ref. \cite{svinin} we found an infinite class of algebraic constraints
for the Volterra lattice (see, below (\ref{1})) but there we do not used
determining equation and showed compatibility in the following equivalent way.
Suppose that one can resolve the equation $h=0$ as 
\begin{equation}
r(i+n)=S_1(r(i),..., r(i+n-1))
\label{boundary1}
\end{equation}
and as
\begin{equation}
r(i-1)=S_2(r(i),..., r(i+n-1)).
\label{boundary2}
\end{equation}
The latter equation can be derived only after shifting $i\rightarrow i-1$.
Identifying $y_1=r(i),..., y_n=r(i+n-1)$ for some value $i=i_0$, one is led
to some first-order 
finite dimensional system of ordinary differential equations on unknown
functions $y_k(t)$. Note that provided that $i$ is some fixed integer,
(\ref{boundary1}) and (\ref{boundary2}) become some boundary conditions for the
Volterra lattice. Then one defines ``new" functions 
$\tilde{y}_1=r(i+1),..., \tilde{y}_n=r(i+n)$. These functions are related with
``old" ones by invertible relations
\begin{equation}
\tilde{y}_1=y_2,...,\; \tilde{y}_{n-1}=y_n,\;\; \tilde{y}_n=S_1(y_1,..., y_n).
\label{maps}
\end{equation}
Then one requires that the collection $\{\tilde{y}_k(t)\}$ also represent
the solution of the finite-dimensional system. It is checked by straightforward
computations.
If so, then one can conclude that algebraic constraint under consideration
is compatible with the Volterra lattice, while the mapping (\ref{maps}) can be
recognized as symmetry transformation for attached finite-dimensional system.

The Letter is organized as follows. In Section 2, we formulate our main theorem
which generalize the result of \cite{svinin}. In Section 3, we write down
finite dimensional systems to which the Volterra lattice is reduced under 
corresponding constraints. Finally, in Section 4, we present relevant 
results for the Toda lattice. Most part of material of this Section can be found
in \cite{svinin}.  

\section{Constraints compatible with Volterra lattice}

Our main result is in the following theorem

{\it Theorem 2.1.}
Each one of the following constraints
\begin{equation}
\sum_{s=1}^{N}r(i+s-1) = \prod_{s=0}^{N+1}r(i+s-1),\;\;
N\ge 1,
\label{1}
\end{equation}
\begin{equation}
\sum_{s=1}^{2M+1}r(i+s-1) = \prod_{s=1}^{M+1}r(i+2s-2),\; M\ge 1
\label{2}
\end{equation}
and
\begin{equation}
\sum_{s=1}^{2M}r(i+s-1)\cdot\sum_{s=1}^{2M}r(i+s)
= \prod_{s=0}^{2M+1}r(i+s-1),\;\;
M\ge 1
\label{f3}
\end{equation}
is consistent with Volterra lattice (\ref{VL}).

To prove the theorem, one need in the following lemma:

{\it Lemma 2.1.}
The quantity
\begin{eqnarray}
I_N(i)&=&\sum_{k=1}^Nr(i+k-1)\left(\sum_{s=k}^Nr(i+s+1)\right) \label{3}\\
      &=&\sum_{k=1}^Nr(i+k+1)\left(\sum_{s=1}^kr(i+s-1)\right)
\label{4}
\end{eqnarray}
is integral for difference system (\ref{1}) (with corresponding $N$).
The quantities $I_{2M}(i)$ and $I_{2M-1}(i)$ are integrals
for the difference system (\ref{2}) and (\ref{f3}), respectively.

Proofs of the above lemma and theorem are quite technical and we find that
it is suitable to put them in Appendix.

\section{Finite-dimensional systems}

Let us present in this Section attached finite-dimensional systems for
all three classes of constraints.

Identify $y_1 = r(i),..., y_{N+1} = r(i+N)$ for some fixed value $i=i_0$.
The constraint (\ref{1}) force this set of functions to be a solution
of the system
\[
\dot{y}_1 = y_1y_2 - \frac{y_1 + ... + y_N}{y_2...y_{N+1}},
\]
\begin{equation}
\dot{y}_k = y_k(y_{k+1} - y_{k-1}),\;\; k =2,..., N,
\label{s1}
\end{equation}
\[
\dot{y}_{N+1} = \frac{y_2 + ... + y_{N+1}}{y_1...y_N} - y_Ny_{N+1}.
\]
From the above theorem we already know that the constraint (\ref{1})
(for any $N$) is compatible with the Volterra lattice. On the level of
the system (\ref{s1}) this means that $r(i)$'s for all $i\in{\bf Z}$ being
expressed via $y_k$'s must solve the Volterra lattice. Consider ``new" variables $\{\tilde{y}_1,..., \tilde{y}_{N+1}\}$
defined by shifting $i\rightarrow i+1$, i.e.
$\tilde{y}_1 = r(i+1),..., \tilde{y}_{N+1} = r(i+N+1)$. Thanks to the 
Theorem 2.1
these ``new" functions also represent a solution of (\ref{s1}) being expressed,
taking into account, (\ref{1}) as
\begin{equation}
\begin{array}{l}
\tilde{y}_1 = y_2,...,\tilde{y}_N = y_{N+1},  \\[0.3cm]
\displaystyle
\tilde{y}_{N+1} =  \frac{y_2 + ... + y_{N+1}}{y_1...y_{N+1}}.
\end{array}
\label{map}
\end{equation}
From what we already know, we can
conclude that any solution of the system (\ref{s1}) supplemented by
the mapping (\ref{map}) gives suitable solution of the Volterra
lattice. Observe that equations yielding $t$-evolution (\ref{s1}) and
(\ref{map}) have common integral
\[
I_N = \sum_{k=1}^{N-1}y_{k+2}\left(\sum_{s=1}^ky_s\right)
+ \frac{(y_1 + ... + y_N)(y_2 + ... + y_{N+1})}{y_1...y_{N+1}}.
\]
Remark that $I_1\equiv 1$.

Similar arguments are relevant for constrains of the second class (\ref{2}).
Corresponding finite-dimensional system together with compatible mapping
(symmetry transformation) read
\[
\dot{z}_1 = z_1\left(
z_2 + \frac{z_1 + ... + z_{2M}}{1 - z_2z_4...z_{2M}}\right),
\]
\begin{equation}
\dot{z}_k = z_k(z_{k+1} - z_{k-1}),\;\; k =2,..., 2M-1,
\label{s2}
\end{equation}
\[
\dot{z}_{2M} = -z_{2M}\left(
\frac{z_1 + ... + z_{2M}}{1 - z_1z_3...z_{2M-1}} + z_{2M-1}\right),
\]
\begin{equation}
\begin{array}{l}
\tilde{z}_1 = z_2,...,
\tilde{z}_{2M-1} = z_{2M}, \\[0.3cm]
\displaystyle
\tilde{z}_{2M} =  \frac{z_1 + ... + z_{2M}}{z_1z_3...z_{2M-1}-1}.
\end{array}
\label{map1}
\end{equation}
Equations (\ref{s2}) and (\ref{map1}) have an integral
\[
\overline{I}_{2M} = \sum_{k=1}^{2M-2}z_{k+2}\left(\sum_{s=1}^kz_s\right)
\]
\[
+ \sum_{k=1}^{2M}z_k\cdot
\frac{z_1z_3...z_{2M-1}\cdot\sum_{k=2}^{2M}z_k +
z_2z_4...z_{2M}\cdot\sum_{k=1}^{2M-1}z_k - \sum_{k=2}^{2M-1}z_k
}{(z_1z_3...z_{2M-1}-1)(z_2z_4...z_{2M}-1)}.
\]

As for constraints 
(\ref{f3}), making of use similar calculations, we obtain the following
system:
\[
\dot{w}_1 = w_1\left(
w_2 + \frac{(w_1 + ... + w_{2M-1})(w_1 + ... + w_{2M})}
{w_1 + ... + w_{2M} - w_1w_2...w_{2M}}\right),
\]
\begin{equation}
\dot{w}_k = w_k(w_{k+1} - w_{k-1}),\;\; k =2,..., 2M-1,
\label{ss2}
\end{equation}
\[
\dot{w}_{2M} = -w_{2M}\left(
\frac{(w_2 + ... + w_{2M})(w_1 + ... + w_{2M})}
{w_1 + ... + w_{2M} - w_1w_2...w_{2M}}+w_{2M-1}\right)
\]
with corresponding symmetry transformation
\begin{equation}
\begin{array}{l}
\tilde{w}_1 = w_2,...,
\tilde{w}_{2M-1} = w_{2M}, \\[0.3cm]
\displaystyle
\tilde{w}_{2M} =  \frac{(w_2 + ... + w_{2M})(w_1 + ... + w_{2M})}
{w_1w_2...w_{2M}- w_1 - ... - w_{2M}}.
\end{array}
\label{map10}
\end{equation}
The system (\ref{ss2}) with compatible mapping (\ref{map10}) has
the integral
\[
\overline{I}_{2M-1} = \sum_{k=1}^{2M-2}w_{k+2}\left(\sum_{s=1}^kw_s\right)
\]
\[
+ \sum_{k=1}^{2M}w_k\cdot
\frac{\sum_{k=1}^{2M-1}w_k\cdot\sum_{k=2}^{2M}w_k}
{w_1w_2...w_{2M}- w_1 - ... - w_{2M}}.
\]

\section{Toda lattice}

Constraints compatible with Toda lattice
\begin{equation}
\begin{array}{l}
\dot{q}_1(i) = q_2(i+1) - q_2(i), \\[0.3cm]
\dot{q}_2(i) = q_2(i)(q_1(i) - q_1(i-1))
\end{array}
\label{toda}
\end{equation}
can be obtained by using well known lattice Miura transformation
\begin{equation}
\begin{array}{l}
q_1(i) = r(2i) + r(2i+1),\\[0.3cm]
q_2(i) = r(2i-1)r(2i).
\end{array}
\label{miura}
\end{equation}
For even $N=2P$ from (\ref{1}) we immediately derive \cite{svinin}
\begin{equation}
\sum_{s=1}^Pq_1(i+s-1) = \prod_{s=1}^{P+1}q_2(i+s-1)
\label{c}
\end{equation}
By analogy with the case of Volterra lattice one can prove

{\it Lemma 4.1.}
The quantity
\begin{eqnarray}
J_P(i)&=&\sum_{k=1}^Pq_1(i+k-1)\left(\sum_{s=k}^Pq_1(i+s)\right)
-\sum_{k=1}^Pq_2(i+k)                                    \nonumber \\
      &=&\sum_{k=1}^Pq_1(i+k)\left(\sum_{s=1}^kq_1(i+s-1)\right)
-\sum_{k=1}^Pq_2(i+k)                                    \nonumber
\end{eqnarray}
is integral for difference system (\ref{c}).

The integral $J_P(i)$ is calculated as  $J_P(i)|_{(\ref{miura})} = I_{2P}(i)$.
Along the lines as in Section 1 we are led to

{\it Theorem 4.1.}
The constraint (\ref{c}) is compatible with Toda lattice (\ref{toda}).

Instead of proving of this Theorem we observe that consistency condition
reads $J_P(i-1)=J_P(i)$ which is valid by virtue of the Lemma 4.1.

To describe the reductions of the Toda lattice in terms of finite-dimensional
systems it is convenient to pass from polynomial to exponential form of the
latter with the help of ansatz
\[
q_1(i) = \dot{u}_i,\;\;
q_2(i) = e^{u_i-u_{i-1}}.
\]
In variables $u_i$ Toda lattice becomes \cite{toda}
\begin{equation}
\ddot{u}_i = e^{u_{i+1}-u_i} - e^{u_i-u_{i-1}}
\label{toda1}
\end{equation}
while the constraint (\ref{c}) turns into
\begin{equation}
\sum_{s=1}^P\dot{u}_{i+s-1} = e^{u_{i+P}-u_{i-1}}
\label{c1}
\end{equation}
Define a finite collection of variables attached to (\ref{c1}) as
$v_1 = u_i,..., v_{P+1} = u_{i+P}$. Then as can be checked the constraint
(\ref{c1}) leads to the system
\[
\ddot{v}_1 = e^{v_2-v_1} -
(\dot{v}_1 + ... + \dot{v}_P)e^{v_{1}-v_{P+1}},
\]
\begin{equation}
\ddot{v}_k = e^{v_{k+1}-v_k} - e^{v_k-v_{k-1}},\;\; k =2,..., P,
\label{s3}
\end{equation}
\[
\ddot{v}_{P+1} =
(\dot{v}_2 + ... + \dot{v}_{P+1})e^{v_{1}-v_{P+1}} - e^{v_{P+1}-v_P}
\]
with corresponding symmetry transformation
\[
\tilde{v}_1 = v_2,...,\;
\tilde{v}_P = v_{P+1},
\]
\[
\tilde{v}_{P+1} = v_1 + \log(\dot{v}_2 + ... + \dot{v}_{P+1}).
\]
Using Miura transformation (\ref{miura}) one can easy prove \cite{svinin}

{\it Proposition 4.1.}
The relations
\[
y_{2k-1}+y_{2k} = \dot{v}_k,\;\; k = 1,..., P
\]
\begin{equation}
y_{2P+1} + \frac{y_2 + ... + y_{2P+1}}{y_1y_2...y_{2P+1}} = \dot{v}_{P+1},
\label{cor}
\end{equation}
\[
y_{2k}y_{2k+1} = e^{v_{k+1}-v_k},\;\; k = 1,..., P
\]
realize the correspondence between the systems (\ref{s3}) and (\ref{s1})
with $N=2P$.

As was noticed in \cite{svinin}, the system (\ref{s3}), for any $P$, admits
Lagrangian and consequently Hamiltonian representation. Lagrangian is given
by
\[
L = \sum_{k < l}\dot{v}_k\dot{v}_l + \sum_{k=1}^Pe^{v_{k+1}-v_k}
+ \left(
\frac{1}{2}\dot{v}_1 + \sum_{k=2}^P\dot{v}_k + \frac{1}{2}\dot{v}_{P+1}
\right)e^{v_1-v_{P+1}}.
\]

It is natural to suppose that systems (\ref{s1}), (\ref{s2}), (\ref{ss2})
and (\ref{s3}) may be integrable in the sense of Liouville-Arnold theorem.
We are going to present the relevant material concerned with first integrals,
Lax pairs, Painlev\'e analysis of the finite-dimensional systems under
consideration in subsequent publications.

\section*{Acknowledgements}

The author is grateful to referees for carefully reading the manuscript 
and for remarks which enabled the presentation of the paper to be improved.

\section*{Appendix}

{\bf A. Proof of the Lemma 2.1.}  

First, notice that it is far from obvious
that $(\ref{3})\equiv (\ref{4})$. To prove this, we need of use
induction by $N$. To this aim, we observe that the recurrence
relation
%\begin{equation}
\[
I_{N+1}(i) = I_{N}(i) + r(i+N+2)\cdot \sum_{s=1}^{N+1}r(i+s-1)
\]
%\label{rec}
%\end{equation}
is valid both for (\ref{3}) and for (\ref{4}) with
$I_1(i)=r(i)r(i+2)$. For $N=1$ the identity $(\ref{3})\equiv
(\ref{4})$ is obvious. Suppose now that this is true for some
positive integer $N$, then
\[
I_{N+1}(i)|_{(\ref{3})} = I_{N}(i)|_{(\ref{3})} + r(i+N+2)\cdot
\sum_{s=1}^{N+1}r(i+s-1) = I_{N}(i)|_{(\ref{4})}
\]
%\begin{equation}
\[
+ r(i+N+2)\cdot \sum_{s=1}^{N+1}r(i+s-1) = I_{N+1}(i)|_{(\ref{4})}
\]
%\end{equation}
Therefore the identity $(\ref{3})\equiv (\ref{4})$ is proved.

Now let us to show that by virtue of (\ref{1}) the relation $I_N(i+1)
= I_N(i)$ is valid. We have
\begin{equation}
I_N(i+1)|_{(\ref{4})} =
\sum_{k=1}^{N-1}r(i+k+2)\left(\sum_{s=1}^kr(i+s)\right) +
r(i+N+2)\cdot \sum_{s=1}^Nr(i+s). \label{have}
\end{equation}
Shifting in (\ref{1}) $i\rightarrow i+2$ one can rewrite it as
$$
r(i+N+2) = \frac{\sum_{s=1}^Nr(i+s+1)} {\prod_{s=0}^Nr(i+s+1)}
$$
Substituting the latter in (\ref{have}) we have
\begin{equation}
I_N(i+1)\stackrel{(\ref{1})}{=}
\sum_{k=1}^{N-1}r(i+k+2)\left(\sum_{s=1}^kr(i+s)\right) +
\frac{\sum_{s=1}^Nr(i+s)\cdot\sum_{s=1}^Nr(i+s+1)}
{\prod_{s=0}^Nr(i+s+1)} \label{again}
\end{equation}
Make of use again the constraint (\ref{1}) in the form
$$
r(i) = \frac{\sum_{s=1}^Nr(i+s)} {\prod_{s=0}^Nr(i+s+1)}.
$$
Substituting that in (\ref{again}) we obtain
$$
I_N(i+1)\stackrel{(\ref{1})}{=}
\sum_{k=1}^{N-1}r(i+k+2)\left(\sum_{s=1}^kr(i+s)\right) +
r(i)\cdot\sum_{s=1}^Nr(i+s-1) = I_N(i).
$$
The similar reasonings are used to prove that $I_{2M}(i)$ is integral
for (\ref{2}) while $I_{2M-1}(i)$ is that for (\ref{f3}).

{\bf B. Proof of the Theorem 2.1.} 

Consider the constraint (\ref{1}). It can be written as
\[
h=\sum_{s=1}^{N}r(i+s-1)-\prod_{s=0}^{N+1}r(i+s-1)=0
\]
By virtue of the Volterra lattice
equations (\ref{VL}) we have the following
\begin{equation}
D_t\left(\sum_{s=1}^{N}r(i+s-1)\right) = r(i+N-1)r(i+N) -
r(i-1)r(i). \label{x}
\end{equation}
On the other hand, taking into account (\ref{1}), we obtain
\begin{equation}
D_t\left(\prod_{s=0}^{N+1}r(i+s-1)\right) =
\sum_{s=1}^{N}r(i+s-1)\cdot(r(i+N) + r(i+N+1) - r(i-1) - r(i-2)).
\label{y}
\end{equation}
To show that this $h$ solves determining equation, one needs to rewrite 
the relation $(\ref{x})=(\ref{y})$ in terms of $I_N(i)$ to use then Lemma 2.1.

Observe that the relation $(\ref{x}) = (\ref{y})$ can be rewritten as
\[
r(i-2)\cdot\sum_{s=1}^{N}r(i+s-1) + r(i-1)\cdot\sum_{s=2}^{N}r(i+s-1)
\]
\begin{equation}
= r(i+N)\cdot\sum_{s=1}^{N-1}r(i+s-1) +
r(i+N-1)\cdot\sum_{s=2}^{N}r(i+s-1) \label{z}
\end{equation}
Adding $I_{N-2}(i)|_{(\ref{3})}$ to l.h.s. of (\ref{z}) and
$I_{N-2}(i)|_{(\ref{4})}$ to r.h.s. of that we obtain
\[
I_{N}(i-2)|_{(\ref{3})} = I_{N}(i)|_{(\ref{4})}.
\]
By virtue of the Lemma 2.1 the latter is identity. Therefore the part of
the Theorem concerning class of constraints (\ref{1}) is proved.
Similar arguments are applied for (\ref{2}) and (\ref{f3}). We only
remark that in these cases we obtain consistency conditions in the
form $I_{2M}(i-1)=I_{2M}(i)$ for (\ref{2}) and
\[
\sum_{s=1}^{2M}r(i+s)\cdot(I_{2M-1}(i) - I_{2M-1}(i-1)) +
\sum_{s=1}^{2M}r(i+s-1)\cdot(I_{2M-1}(i+1) - I_{2M-1}(i)) = 0.
\]
for (\ref{f3}), respectively.


\begin{thebibliography}{99}

\bibitem{volterra} V. Volterra, Le\c{c}ons sur la th\'eorie mathematique
de la lutte sur la vie (Gautier-Villars, Paris, 1931).

\bibitem{zakharov1} V.E. Zakharov, S.L. Musher, A.M. Rubenchik, JETP Lett.
19 (1974) 151.

\bibitem{zakharov2} V.E. Zakharov, S.L. Musher, A.M. Rubenchik, Sov. Phys.-JETP
42 (1976) 80.

\bibitem{manakov} S.V. Manakov, Sov. Phys.-JETP. 40 (1975) 269.

\bibitem{veselov} A.P. Veselov, Teor. Mat. Fiz.  (1987) 154.
(in Russian)

\bibitem{ver} V.L. Vereshchagin, Teor. Mat. Fiz. 111 (1997) 335.
(in Russian)

\bibitem{bog} O.I. Bogoyavlenskii, Math. USSR-Izv. 31 (1988) 435. 

\bibitem{gurel} B. G\"urel, M. G\"urses, I. Habibullin, J. Math. Phys.
36 (1995) 6809.

\bibitem{h1}
I.T. Habibullin, Phys. Lett. A 207 (1995) 263.

\bibitem{h2}
V.E. Adler, I.T. Habibullin, J. Phys. A: Math. Gen. 28 (1995) 6717.

\bibitem{dam1} P.A. Damianou, R.L. Fernandes, Rep. Math. Phys. 50 (2002) 361.

\bibitem{dam2} P.A. Damianou, S.P. Kouzaris, Physica D. 195 (2004) 50.

\bibitem{sidorov}
A.F. Sidorov, V.P. Shapeev, N.N. Yanenko, Method of differential constraints
and its applications in gas dynamics (Nauka, Novosibirsk, 1984), (in Russian).

\bibitem{svinin} A.K. Svinin, Theor. Math. Phys. 124 (2000) 1211.

\bibitem{toda} M. Toda, Progr. Theor. Phys. Suppl. 59 (1976) 1.

\end{thebibliography}
\end{document}